\RequirePackage{fix-cm}
\documentclass[smallcondensed,final]{svjour3}     
\smartqed  
\usepackage{graphicx}
\usepackage{latexsym}
\journalname{Hyperfine Interactions}
\begin{document}
\title{Physics topics at KLOE-2
}

\author{M. Silarski on behalf of the KLOE-2 Collaboration
\thanks{The KLOE- 2 Collaboration: F.~Archilli, D.~Babusci,
D.~Badoni, I.~Balwierz, G.~Bencivenni, C.~Bini, C.~Bloise,
V.~Bocci, F.~Bossi, P.~Branchini, A.~Budano, S.~A.~Bulychjev,
L.~Caldeira~Balkest\aa hl, P.~Campana, G.~Capon, F.~Ceradini,
P.~Ciambrone, E.~Czerwi\'nski, E.~Dan\'e, E.~De~Lucia,
G.~De~Robertis, A.~De~Santis, G.~De~Zorzi, A.~Di~Domenico,
C.~Di~Donato, D.~Domenici, O.~Erriquez, G.~Fanizzi, G.~Felici,
S.~Fiore, P.~Franzini, P.~Gauzzi, G.~Giardina, S.~Giovannella,
F.~Gonnella E.~Graziani, F.~Happacher, B.~H\"oistad,
L.~Iafolla, E.~Iarocci, M.~Jacewicz, T.~Johansson,
A.~Kowalewska, V.~Kulikov, A.~Kupsc, J.~Lee-Franzini, F.~Loddo,
G.~Mandaglio, M.~Martemianov, M.~Martini, M.~Mascolo,
M.~Matsyuk, R.~Messi, S.~Miscetti, G.~Morello, D.~Moricciani,
P.~Moskal, F.~Nguyen, A.~Passeri, V.~Patera, I.~Prado~Longhi,
A.~Ranieri, C.~F.~Redmer, P.~Santangelo, I.~Sarra, I.~Sarra,
B.~Sciascia, A.~Sciubba, M.~Silarski, C.~Taccini, L.~Tortora,
G.~Venanzoni, R.~Versaci, W.~Wi\'slicki, M.~Wolke, J.~Zdebik} }
\institute{M. Silarski \at Institute of Physics, Jagiellonian
University,PL-30-059 Cracow, Poland
\\
              Tel.: +48-12-6635533; Fax: +48-12-6342038\\
              \email{Michal.Silarski@lnf.infn.it}           
}
\date{Received: date / Accepted: date}

\maketitle

\begin{abstract}
The goal of the KLOE--2 experiment operating at the upgraded
DA$\Phi$NE $e^+e^-$ collider is to collect an integrated
luminosity of about 20 fb$^{-1}$ over 3-4 years of running. The
KLOE--2 apparatus is now equipped with an inner tracker, new
scintillation calorimeters and tagging detectors for $\gamma
\gamma$ physics. These will allow measurements to refine and
extend the KLOE programme on kaon physics and tests of
fundamental symmetries as well as quantum interferometry. Here
the latest results from the KLOE data analysis are presented
and the perspectives at \mbox{KLOE-2} outlined.%
\end{abstract}
\section{Introduction}
\label{intro} In the 2000--2006 data-taking period the KLOE
detector, operating at the DA$\Phi$NE accelerator in the
Laboratori Nazionali di Frascati, acquired a total integrated
luminosity of 2.5~fb$^{-1}$ at the $\phi$ mass peak
(corresponding to about $10^{10}$ $\phi$ decays) and
250~pb$^{-1}$ at $\sqrt{s}=1$~GeV (giving about 10$^8$ $\eta$
mesons). This allowed precise studies to be carried out on
charged and neutral kaon physics, low energy QCD, as well as
tests of CP and CPT conservation~\cite{kloe2008}. In 2008 the
Accelerator Division of the Frascati Laboratory tested a new
interaction scheme to allow the beam size to be reduced and the
luminosity increased. The test was successful and presently
DA$\Phi$NE can reach a peak luminosity of $5\times
10^{32}$~cm$^{-2}$s$^{-1}$, which is a factor of three better
than previously obtained~\cite{pantaleo}. Following these
achievements, the data-taking campaign using the upgraded KLOE
detector on the improved machine will start in Autumn 2011. The
goal is to collect an integrated luminosity of about
20~fb$^{-1}$ over 3-4 years of running.
%
%
\section{The KLOE experiment at DA$\Phi$NE}
\label{KLOE:DAFNE}
DA$\Phi$NE is a $e^+e^-$ collider operating near the $\phi$
meson mass peak, at a centre-of-mass energy $\sqrt{s} =
1019.45$~MeV~\cite{kloe2008}. The heart of DA$\Phi$NE consists
of two rings in which 120 bunches of electrons and positrons
are stored. Electrons are accelerated in the Linac, stored and
cooled in the accumulator, and then transferred as a single
bunch to the ring. Positrons are first created at an
intermediate station in the linac using 250~MeV electrons and
then follow the same procedure as for electrons.
Electrons and positrons collide with small transverse momenta
and produce $\phi$ mesons almost at rest ($\beta_{\phi} \approx
0.015$). The $\phi$ decay mainly into $K^+K^-$ (49\%), $K_SK_L$
(34\%), $\rho\pi$ (15\%) and $\eta\gamma$ (1.3\%). The decay
products are registered using the KLOE detection setup. This
consists of a $\approx 3.3$~m long cylindrical drift chamber,
with a diameter of $\approx 4$~m, which is surrounded by the
electromagnetic calorimeter. The detectors are placed in the
axial magnetic field of a superconducting solenoid of strength
$B=0.52$~T.
The KLOE drift chamber is constructed out of a carbon fibre
composite with low-Z and low density and uses a gas mixture of
helium~(90\%) and isobutane (10\%). It provides tracking in
three dimensions, with a resolution in the transverse plane of
about 200~$\mu$m, a resolution in the $z$-coordinate
measurement of about 2~mm and 1~mm in the decay vertex
position. The momentum of a particle is determined from the
curvature of its trajectory in the magnetic field with a
fractional accuracy $\sigma_p/p=0.4\%$ for polar angles greater
than $45^{\circ}$~\cite{kloe2008}.
The KLOE electromagnetic calorimeter consists of a barrel built
out of 24 trapezoidally shaped modules and side detectors (so
called endcaps) read out from both sides by a set of
photomultipliers~\cite{kloe2008}. Each of the modules is constructed
out of 1 mm scintillating fibres embedded in 0.5 mm lead foils
to speed up the showering processes. This detector allows
measurements of particle energies and flight times with
accuracies of $\sigma_E=5.7\%E/\sqrt{E[{\rm GeV}]}$ and
$\sigma(t)=57~\textrm{ps}/\sqrt{E[{\rm
GeV}]}~\oplus~140~\textrm{ps}$, respectively. Analysis of the
signal amplitude distributions allows one to determine the
place where the particle hit the calorimeter module with
accuracy of about 1~cm in the plane transverse to the fibre
direction. The longitudinal coordinate precision is energy
dependent: $\sigma_z=1.2~\textrm{cm}/\sqrt{E[{\rm GeV}]}$.
Since the $\phi$ mesons are produced almost at rest, kaons
arising from the decay move at low speed with their relative
angle being close to 180$^{\circ}$. As a consequence, the decay
products are registered in well separated areas of the
detector, which allows identification of $K_L$ mesons using
reconstructed decays of $K_S$ (so called $K_L$ tagging) and
vice versa. This is a special feature of the DA$\Phi$NE
accelerator which, together with the KLOE detector, is a unique
laboratory for kaon physics~\cite{epjC}.
\section{From KLOE to KLOE--2}
In recent years, a new scheme for the machine based, on
Crab-waist optics and a large Piwinsky angle~\cite{pantaleo},
has been proposed to increase the DA$\Phi$NE luminosity. This
has been tested successfully and it has motivated the start of
a new KLOE run with an improved detector setup, named
\mbox{KLOE-2}. This aims to complete the KLOE physics programme
and perform a new set of interesting measurements~\cite{epjC}.
For the forthcoming run we have improved the performance of
KLOE by adding new sub-detector systems: the tagger system for
$\gamma\gamma$ physics studies, the Inner Tracker based on the
Cylindrical GEM technology, a tile calorimeter surrounding the
inner quadrupoles (QCALT), and a calorimeter between the
interaction point (IP) and the first inner quadrupole (CCALT).
The tagging system is made up of two different detectors which
are already installed and ready for data taking. The Low Energy
Tagger (LET) is a small calorimeter placed inside KLOE near the
IP, consisting of LYSO crystals read out by silicon
photomultipliers. This sub-detector will serve to measure
electrons and positrons from $\gamma\gamma$ interactions within
a wide energy range centred around 200~MeV with an accuracy
$\sigma_E\sim10$\%. The second tagger, which is called the High
Energy Tagger (HET), provides a measurement of the displacement
of the scattered leptons with respect to the main orbit.
This position detector consists of 30 small BC418 scintillators
$3\times3\times5$~mm$^3$, which provide a spatial resolution of
2~mm (corresponding to a momentum resolution of $\sim 1$~MeV/$c$).
The output light is collected by light guides with SiPM
sensors.
The HET allows measurements of particle energies with
an accuracy of $\sigma_E \sim 2.5$~MeV and time with a
resolution of $\sigma_t\sim 200$~ps. 
To improve the acceptance for low momentum tracks, and the
vertex reconstruction near the interaction point, we are
building the inner tracking chamber. This employs a novel
technology with cylindrical GEM (Gas Electron Multiplier)
detectors. It will be composed of four concentric layers that
will provide 2-D points on a cylinder of known radius. Each
layer is a triple-GEM chamber with cathode and anode made of
thin polyamide foils.
We will also install two additional calorimeters, named QCALT
and CCALT. QCALT will be a 1~m long dodecagonal structure
covering the region of the new quadrupoles. It is composed of a
sampling of five layers of 5~mm thick scintillator plates
alternated with 3.5~mm thick tungsten plates, for a total depth
of 4.75~cm.
The crystal calorimeter CCALT will cover the low  polar angle
region to increase the acceptance for very forward photons down
to $8^{\circ}$. The basic layout consists of two small barrels
of LYSO crystals that are read out with APD photosensors. A
timing resolution between 300 and 500~ps is expected for 20~MeV
photons.
\section{Recent KLOE results and the ongoing analysis in kaon physics}
In addition to the preparation for the first KLOE-2 data-taking
period (so-called Step0) and the activities for the design and
construction of the new sub-detectors, there are still several
ongoing physics analyses of the KLOE data. In kaon physics,
apart from the recently published results of a precise $K_S$
lifetime measurement~\cite{ts}, there are several studies of
neutral kaon interferometry and rare kaon decays, for example
$K_{S}K_{L} \to \pi^+\pi^-\pi^+\pi^-$ interferometry or the
$K_S \to \pi^0\pi^0\pi^0$ branching ratio measurement. These
will be described briefly here.
The decay of $K_S$ meson into three pions has not yet been
observed, and the best limit on the branching ratio
$BR(K_{s}\rightarrow 3\pi^0) < 1.2\times10^{-7}$~\cite{Matteo}
is about two orders of magnitude larger than predictions based
on the Standard Model. Moreover, this process violates $CP$
symmetry and, assuming $CPT$ invariance, allows one to
investigate direct $CP$ violation. At KLOE this decay is
reconstructed by searching for events with a $K_L$ interaction
in the calorimeter (so called $K_L$--crash), six photon
clusters and no tracks from the interaction point. The
background originates mainly from $K_S \to 2\pi^0$ events with
two spurious clusters from splittings or accidental activity or
is due to false $K_L$--crash tags from $\phi \to K_{S}K_{L} \to
\pi^+\pi^-\pi^0\pi^0\pi^0$ events. In the latter case, charged pions
from $K_S$ decays interact in the low-beta insertion
quadrupoles, ultimately simulating the $K_L$--crash signal,
while $K_L$ decays close to the interaction point produce six
photons~\cite{Matteo}. To reduce the background, we first
perform a kinematic fit with 11 constraints: energy and
momentum conservation, the kaon mass and the velocities of the
six photons.
\begin{figure*}
\centering
\includegraphics[width=0.4112\textwidth]{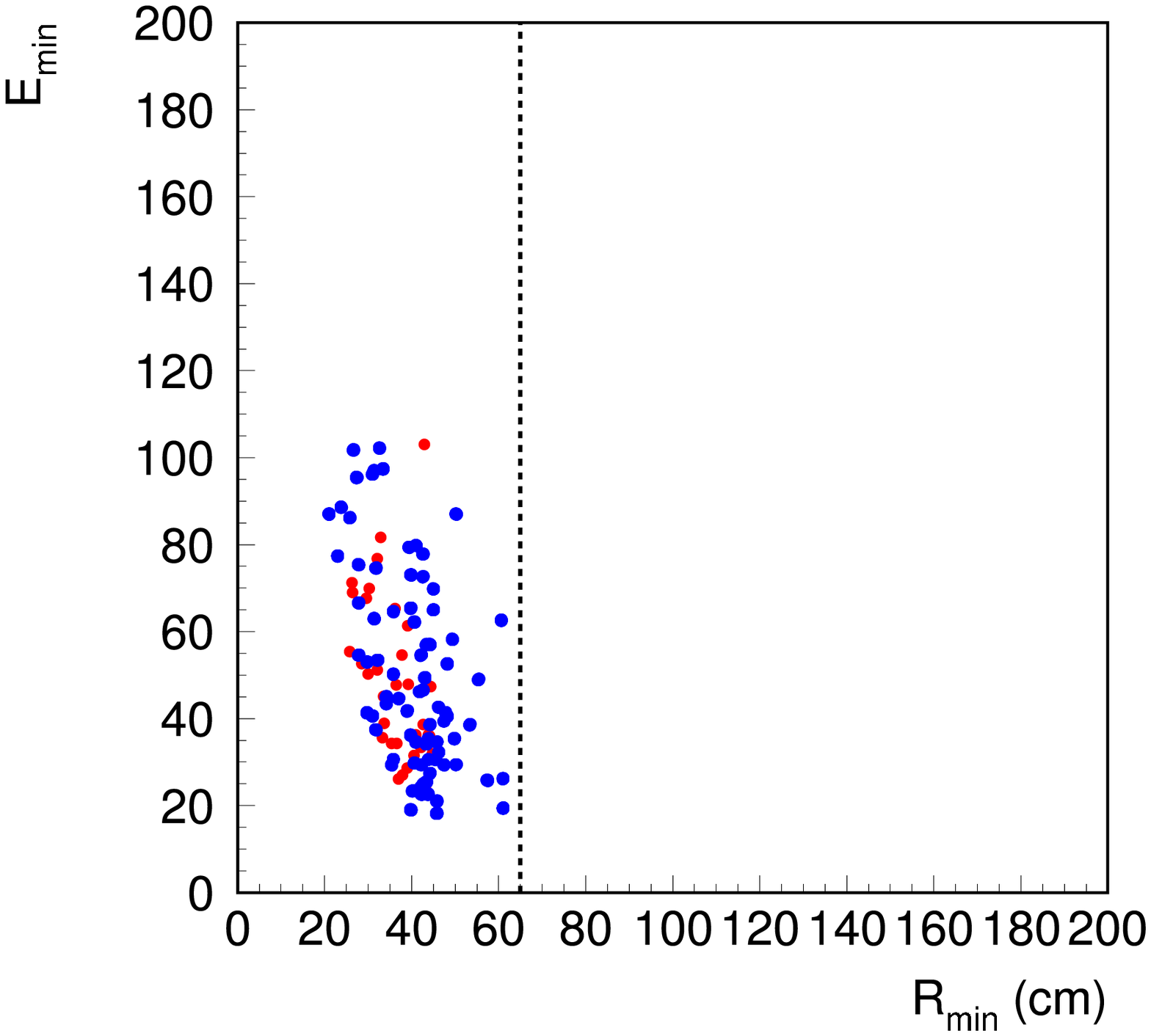}
\includegraphics[width=0.4112\textwidth]{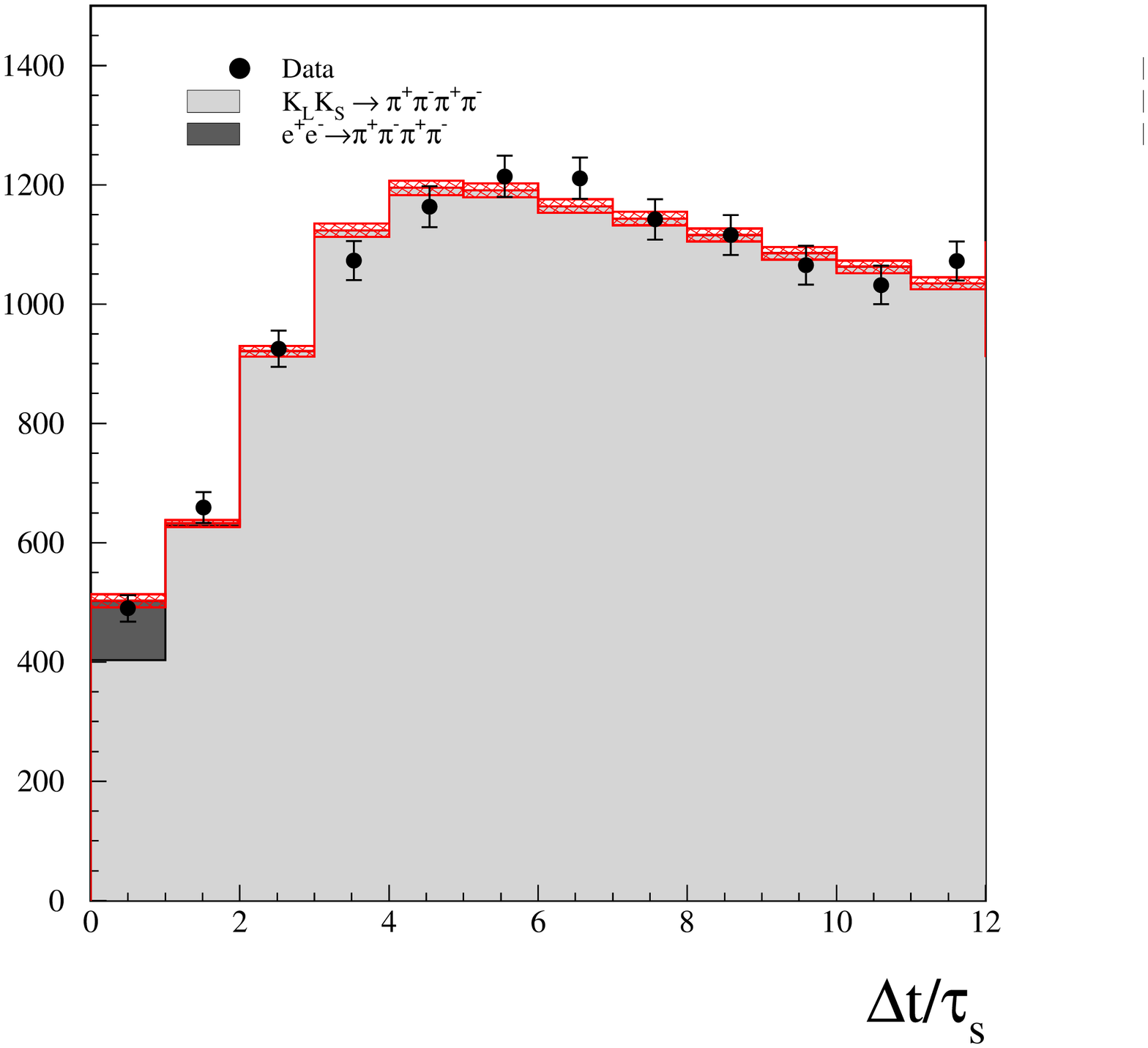}
\caption{ Left panel: The distribution of the minimal energy of the
cluster versus minimal distance ($R_{\rm min}$) between
clusters in the event for data (red) and MC (blue). The dashed
line corresponds to the $R_{\rm min}$ cut used. Right panel:
Number of events as a function of the difference in the decay
time of the two $\pi^+\pi^-$ vertices in $K_S$ lifetime units.
The measured $I(\pi^+\pi^-, \pi^+\pi^-,\Delta t)$ distribution
is fitted: black points are data and the fit result is shown by
the histogram.} \label{ks3pi0}
\end{figure*}
Cutting at a reasonable $\chi^2$ value reduces considerably the
background from false $K_L$--crash events with little signal
loss. In order to gain a good rejection of background from
events with split or accidental clusters, we look at the
correlation between the following two $\chi^2$--like
discriminating variables, $\chi^{2}_{3\pi}$ and
$\chi^{2}_{2\pi}$. $\chi^{2}_{3\pi}$ is the quadratic sum of
the residuals between the nominal $\pi^0$ mass and the
invariant masses of three photon pairs formed from the six
clusters present. $\chi^{2}_{2\pi}$ is based instead on energy
and momentum conservation in the $\phi \to K_SK_L, K_S \to
\pi^0\pi^0$ decay hypothesis, as well as on the invariant
masses of two photon pairs. Both variables are evaluated with
the most favorable cluster pairing in each case~\cite{Matteo}.
In addition, in order to improve the quality of the photon
selection using $\chi^{2}_{2\pi}$, we cut on the variable
$\Delta E=(m_{\Phi}c^2/2 - \sum E_{\gamma_{i}})/\sigma_{E}$, where
$\gamma_i$ is the i--$th$ photon from the four chosen in the
$\chi^{2}_{2\pi}$ estimator and $\sigma_E$ is the appropriate
resolution. For $K_S \to 2\pi^0$ decays plus two background
clusters, we expect $\Delta E\sim 0$ while, for $K_S \to
3\pi^0$, $\Delta E \sim m_{\pi^0}c^2/\sigma_E$. At the end of the
analysis we cut also on the minimal distance between the photon
clusters to refine the rejection of events with split clusters
(see left panel of Fig.~\ref{ks3pi0}).
Applying the preliminary selection cuts, we find zero candidates in
1.7~fb$^{-1}$ of real data with zero events expected from Monte
Carlo, corresponding to an effective statistics of two times
that of the data. This results in a new preliminary upper limit
on the branching ratio $BR(K_S \to 3\pi^0) < 2.9 \times 10^{-8}$,
which suggests that a first observation of the decay might be
feasible at KLOE-2.
A unique feature of the $\Phi$-factory is the production of
neutral kaon pairs in a pure quantum state so that we can study
quantum interference effects and tag pure monochromatic $K_S$
and $K_L$ beams. The decay rate of the system, for example, to
the $\pi^+\pi^-\pi^+\pi^-$ final state is proportional to:
\begin{eqnarray}
I(\pi^+\pi^-, \pi^+\pi^-,\Delta t)~\propto~e^{-\Gamma_L \Delta t}~+~e^{-\Gamma_S \Delta t}
-2e^{-\frac{\Gamma_L~+~\Gamma_S}{2}\Delta t}\cos(\Delta m \Delta t)~,
\label{inter}
\end{eqnarray}
where $\Delta t$ is the time difference between the decays of
the two kaons. If the neutral kaon system evolves in time as a
pure quantum state, kaons cannot decay at the same time due to the
destructive interference. But there are several potential
mechanisms leading to decoherence of the state which may result
from fundamental modifications of Quantum Mechanics or CPT
violation induced, e.g.,by quantum gravity~\cite{epjC}. Thus, by
measuring the $\Delta t$ distribution, we can test the foundations
of Quantum Mechanics as well as different phenomenological
models of Quantum Gravity.
At KLOE, the selection of a $\phi \to K_SK_L \to \pi^+\pi^-\pi^+\pi^-$ signal
requires two vertices, each with two opposite-curvature tracks inside the
drift chamber, with invariant mass and total momentum compatible with two
neutral kaon decays. The resolution in $\Delta t$, the absolute value of the
time difference of two $\pi^+\pi^-$ decays, benefits from the precise
momentum measurements and from the completeness of the kinematics of the
events~\cite{CPT2}. The experimental $\Delta t$ distribution is fitted with
Eq.~(\ref{inter}), modified by parameters expressing decoherence in the
different models described in~\cite{epjC}. The fit is performed taking into
account the resolution and detection efficiency, the background from coherent
and incoherent $K_S$ regeneration on the beam pipe wall, and the small
contamination from the non-resonant $e^+e^- \to \pi^+\pi^-\pi^+\pi^-$
channel.
The resulting distributions  are shown together with the fit results in
Fig.~\ref{ks3pi0}. On the basis of the 1.7~fb$^{-1}$ of data, useful
estimates could be made of several decoherence and CPT violating parameters.
The details of the analysis and the results obtained can be found in
Ref.~\cite{CPT}. It is worth mentioning that we have almost finished a
refined analysis of the KLOE data with several improvements in the
methodology. Moreover, KLOE-2 with increased statistics ($\times 4$ in Step0)
and improved resolution in the vertex reconstruction ($\times 3$ with the
Inner Tracker) will be able to achieve the best experimental sensitivity in
some observables, hopefully reaching the level of the Planck
scale~\cite{epjC}.
\section{Summary}
KLOE is a high precision experiment which allows detailed studies of both
kaons and light scalar mesons, as well as tests of the conservation of CP,
CPT and low energy QCD. The success of the DA$\Phi$NE upgrade motivated a new
experiment, KLOE-2, which aims at completing and extending the KLOE physics
programme. We have started design and construction of new sub-detectors,
which will improve the detection performance. The tagging system for
$\gamma\gamma$ physics is installed and ready for the first phase of the
experiment in which we expect to accumulate about 5~fb$^{-1}$. The next
data-taking campaign during 2013-15 will be conducted with the Inner Tracker
and improved photon acceptance brought about by the calorimeters in the final
focusing region. The total integrated luminosity expected in this second
phase is about 20~fb$^{-1}$.
\begin{acknowledgements}
The author would like to express his gratitude to prof. Colin Wilkin for proof
reading of the article and many useful comments.\\
We acknowledge support by Polish Ministry of Science and Higher Education through the Grant
No. 0469/B/H03/2009/37, and by the European Community-Research Infrastructure Integrating Activity
``Study of Strongly Interacting Matter`` (acronym HadronPhysics2, Grant Agreement n. 227431)
under the Seventh Framework Programme of EU.
 
\end{acknowledgements}

\end{document}